\documentclass[epj]{svjour}
\usepackage{amssymb}
\usepackage{amsmath}
\usepackage{graphics}
\begin{document}
\title{Teleportation of the one-qubit state with environment-disturbed recovery operations}

\author{M.L. Hu\thanks
{e-mail: mingliang0301@xupt.edu.cn}
}
\institute{School of Science, Xi'an University of Posts and
           Telecommunications, Xi'an 710061, P.R. China}
\date{Received: date / Revised version: date}
%
\abstract{ We study standard protocol $\mathcal{P}_0$ for
teleporting the one-qubit state with both the transmission process
of the two qubits constitute the quantum channel and the recovery
operations performed by Bob disturbed by the decohering environment.
The results revealed that Bob's imperfect operations do not
eliminate the possibility of nonclassical teleportation fidelity
provided he shares an ideal channel state with Alice, while the
transmission process is constrained by a critical time $t_{0,c}$
longer than which will result in failure of $\mathcal{P}_0$ if the
two qubits are corrupted by the decohering environment. Moreover, we
found that under the condition of the same decoherence rate
$\gamma$, the teleportation protocol is significantly more fragile
when it is executed under the influence of the noisy environment
than those under the influence of the dissipative and dephasing
environments.
} 
\maketitle
\section{Introduction}
\label{intro} Quantum teleportation \cite{Ref1}, the disembodied
transport of a quantum state based on the nonlocal properties of an
entangled state resource, has been demonstrated to be one of the
most peculiar and fascinating aspects of quantum information theory.
Together with the help of local operations and classical
communication (LOCC), it allows sending the quantum information from
a sender, conventionally named Alice, to a distant receiver Bob,
with fidelity (see Section 2) better than that achievable via
classical communication alone, with the cost of destroying the
original state. Due to the important role it played in quantum
information theory, a lot of theoretical works
\cite{Ref2,Ref3,Ref4,Ref5} have been devoted to it in recent years.
Experimental realization of quantum teleportation has also been
successfully demonstrated with photonic qubits \cite{Ref6,Ref7} and
atomic qubits \cite{Ref8,Ref9}.

The practical implementation of quantum teleportation begins with
the preparation of a pair of entangled qubits which are shared by
two parties, Alice and Bob. This step establishes a quantum link
between them (see Figure 1). Alice receives a state to be teleported
and performs the Bell state measurement on her two qubits, and then
communicates classically the measurement result (two bits of
classical information) to Bob, who uses it to perform recovery
operations on his qubit, thus completing the process of
teleportation. The perfect implementation of this quantum protocol
requires the sharing of the maximally entangled channel state and
complete coherent control over a system's quantum state. In real
circumstances, however, decoherence due to the inevitable
interaction of the system with the surrounding environment makes it
very difficult to prepare the maximally entangled channel states
\cite{Ref10,Ref11,Ref12,Ref13}, and the amount of entanglement may
be further reduced when the two qubits being distributed to Alice
and Bob because during the transmission process, the qubits may also
be exposed to decohering environment. For this reason, a number of
schemes using non-maximally entangled state as resource have been
proposed \cite{Ref2,Ref3,Ref4}. These works reveal several
interesting aspects of quantum entanglement in terms of their
teleportation capacity. Particularly, it is shown in reference
\cite{Ref3} that standard teleportation with an arbitrary entangled
mixed state resource is equivalent to a generalized depolarizing
channel with probabilities given by the maximally entangled
components of the resource.
\begin{figure}
\centering
\resizebox{0.40\textwidth}{!}{%
\includegraphics{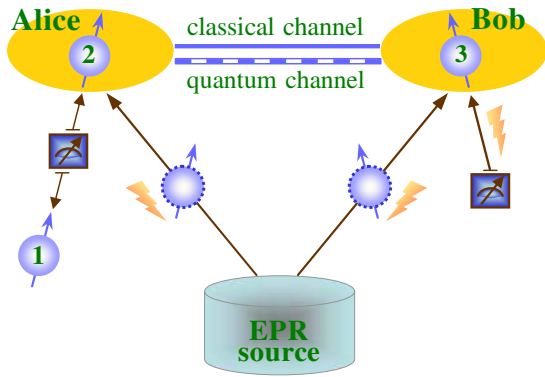}}
\caption{(Color online) Schematic picture of quantum teleportation
of the one-qubit state (encoded at qubit 1) in the decohering
environment. Here the ``ammeter" symbols represent quantum
measurements, while the ``lightning" symbols represent decoherence
imposed by the external environment. Moreover, we assume that the
decoherence only takes place when the two qubits (2 and 3) being
transmitted form the EPR source to Alice and Bob (using time, say
$t_0$) and when Bob performs the recovery operations.} \label{fig:1}
\end{figure}

Although teleportation with system decoherence have been studied by
many authors in recent years \cite{Ref14,Ref15,Ref16,Ref17,Ref18},
we noted that the effects of imperfect operations (e.g., the Bell
state measurement performed by Alice and the recovery operations
performed by Bob) on quantum teleportation has seldom been
considered. However, a through understanding of this problem is
obviously vital for the achievement of high efficient and long
distance quantum communication, for in the presence of a decohering
environment, it is difficult to execute the complete coherent
control of a quantum state. Indeed, several recent works
\cite{Ref14,Ref18} have demonstrated that the noisy operations may
have significant influence on reducing fidelity of the expected
outcomes. Stimulated by this observation, in the present paper we
would like to reexamine the standard teleportation protocol
$\mathcal{P}_0$ \cite{Ref1} with the addition of
environment-disturbed recovery operations. Although this issue is
somewhat similar to that discussed in reference \cite{Ref18}, we
concentrated on, however, different mechanisms of decoherence (see
Section 2 for more detail), and thus one may expects that it will
include new features characteristic of the considered system here.
Our results revealed that in order to execute the quantum
teleportation protocol with fidelity better than the classical
communication alone will do, the rotation rate $\omega$ of Bob's
recovery operations must be larger than a critical value. Moreover,
if the channel state is prepared maximally and distributed to Alice
and Bob without decoherence, then Bob's imperfect recovery
operations do not rule out the possibility of nonclassical
teleportation of a quantum state. For the decohered channel states,
however, the transmission time $t_0$ (see Figure 1) of the two
qubits must be shorter than a critical value to ensure success of
the standard teleportation protocol.

\section{Basic formalism}
\label{sec:2} During the teleportation process, the unavoidable
interaction of an open quantum system with its surrounding
environment is an important source of decoherence \cite{Ref19}. In
order to describe such process, a master equation approach can be
used. Under the assumption of Markovian and Born approximations and
after performing the partial trace over the environmental degrees of
freedom, the reduced density operator $\rho$ of the open quantum
system evolves according to a general master equation in the
Lindblad form \cite{Ref19,Ref20}
\begin{equation}
 {d{\rho}\over{dt}}=-\text{i}[\hat{H},\rho]+{\gamma\over 2}
 \sum_{k,i}(2\mathcal{L}_{k,i}\rho\mathcal{L}_{k,i}^\dag-
 \mathcal{L}_{k,i}^\dag\mathcal{L}_{k,i}\rho-
 \rho\mathcal{L}_{k,i}^\dag\mathcal{L}_{k,i}),
\end{equation}
where $\hat{H}$ denotes the Hamiltonian of the system, and $\gamma$
is the phenomenological parameter that describes the coupling
strengths of the qubits with their respective environment. The
generators of decoherence here are defined in terms of the raising
and lowering operators $\sigma^{\pm}=(\sigma^1\pm
\text{i}\sigma^2)/2$ ($\sigma^n$ with $n=0,1,2,3$ signify the
$2\times 2$ identity matrix and the three Pauli spin operators) as
$\mathcal{L}_{k}=\sigma_{k}^{-}$ for the dissipative environment,
$\mathcal{L}_{k,1}=\sigma_{k}^{-}$ and
$\mathcal{L}_{k,2}=\sigma_{k}^{+}$ for the noisy environment, and
$\mathcal{L}_{k}=\sigma_{k}^{+}\sigma_{k}^{-}$ for the dephasing
environment \cite{Ref10}. Moreover, we have assumed that during the
decoherence process each qubit of the open system interacts only,
and independently, with its own environment. This assumption is
reasonable provided the constituents composing the quantum system
are separated by distances large enough \cite{Ref19}.

In the present work, we explore standard teleportation protocol of
the one-qubit state when it is executed in the presence of
dissipative, noisy and dephasing environments \cite{Ref10}. For
simplicity, we consider throughout this paper the situation in which
Alice's Bell state measurement is perfect, while the decoherence
only takes place during the establishment of the channel state
(e.g., a third party prepares the maximally entangled Bell state at
time $t=0$, and then sends one qubit to Alice and another one to Bob
after a time interval $t_0$. During the transmission process, the
two qubits may be exposed to the decohering environment, and thus
degrades entanglement between them) as well as Bob performs the
recovery operations (see Figure 1 for an illustration of this
process).

For the ideal situation (i.e., no decoherence), if Alice and Bob
share one of the maximally entangled Bell state
$|\rm{\Psi}^{0}\rangle=(|00\rangle+|11\rangle)/\sqrt{2}$, then
according to the definition of the standard teleportation protocol
$\mathcal{P}_0$ as stated by Bennett et al. \cite{Ref1}, the joint
state composed of the state to be teleported and the quantum channel
can be expressed as
\begin{equation}
 \rho_{123}={1\over
 4}\sum_{m=0}^{3}{\rm{\Pi}}_{12}^{m}(\sigma^{m}\rho_{\text{in}}\sigma^{m}),
\end{equation}
where
$\rho_{\text{in}}=|\varphi_{\text{in}}\rangle\langle\varphi_{\text{in}}|$,
and $|\varphi_{\text{in}}\rangle$ is the unknown one-qubit state
Alice seeks to teleport to Bob, which can be represented on a Bloch
sphere as $|\varphi_{\text{in}}\rangle=\cos(\theta/2)|0\rangle+
e^{\text{i}\phi}\sin(\theta/2)|1\rangle$, where
$0\leqslant\theta\leqslant\pi$ and $0\leqslant\phi\leqslant 2\pi$
are the polar and azimuthal angles, respectively. Moreover,
${\rm{\Pi}}_{12}^{m}=|{\rm{\Psi}}^{m}\rangle\langle{\rm{\Psi}}^{m}|$
($m=0,1,2,3$) denote the Bell state measurements performed by Alice,
with $|\rm{\Psi}^{0,3}\rangle=(|00\rangle\pm|11\rangle)/\sqrt{2}$
and $|\rm{\Psi}^{1,2}\rangle=(|01\rangle\pm|10\rangle)/\sqrt{2}$
being the four Bell states. It follows immediately from equation (2)
that if Alice's measurement outcome is $m$, then Bob's unitary
operation to recover $|\varphi_{\text{in}}\rangle$ will be
$\sigma^{m}$. For general cases, the channel state $\rho^{(\alpha)}$
established between Alice and Bob at an arbitrary time $t_0$ will be
mixed due to the unavoidable interaction of the two-qubit system
with the decohering environment, and thus severely undermines the
feasibility of entanglement as a resource for teleportation. The
explicit forms of $\rho^{(\alpha)}$ can be obtained by solving the
appropriate master equation (1) with
$\rho^{(\alpha)}(0)=|\rm{\Psi}^{0}\rangle\langle\rm{\Psi}^{0}|$ as
the initial condition. Here $\alpha=p$, $di$, $no$ or $de$ indicates
the case that the channel state is protected perfectly, or corrupted
by the dissipative, noisy or dephasing environment. If Alice's
measurement has outcome $m$, she tells this measurement result to
Bob in a classical way, then the output state (teleported via the
state $\rho^{(\alpha)}$) after Bob's recovery operations conditioned
on the two bits of classical information received from Alice is
given by
\begin{equation}
 \mathcal{E}_{m}^{(\beta)}[\rho^{(\alpha)}]={1\over{P_{m}^{(\alpha)}}}
 \mathcal{R}_{m}^{(\beta)}\{\text{tr}_{1,2}[({\rm{\Pi}}_{12}^{m}\otimes\sigma_{3}^0)
 (\rho_{\text{in}}\otimes\rho^{(\alpha)})]\},
\end{equation}
where
$P_{m}^{(\alpha)}=\text{tr}_{1,2,3}[({\rm{\Pi}}_{12}^{m}\otimes\sigma_{3}^0)
(\rho_{\text{in}}\otimes\rho^{(\alpha)})]$ is the probability for
Alice to get the measurement outcome $m$. $\sigma_{3}^0$ is the
$2\times 2$ identity matrix acting on qubit 3.
$\mathcal{R}_m^{(\beta)}$ is a trace-preserving quantum operation
carried out by Bob for the purpose of accomplishing the
teleportation process, where $\beta=di$, $no$ or $de$ indicates if
Bob's operation is infected with the dissipative, noisy or dephasing
environment. The explicit form of $\mathcal{R}_m^{(\beta)}\{\rho\}$
can be derived from equation (1) with the system Hamiltonian
$\hat{H}=\hat{H}_{m}=-\omega\sigma^{m}/2$, which generates an
anticlockwise coherent rotation of a qubit about the $m$-axis at the
rate $\omega$ \cite{Ref20}.

The resemblance of the two quantum states before and after
teleportation can be quantified by the fidelity
$f_m^{(\beta)}[\rho^{(\alpha)}]=\langle\varphi_{\text{in}}|\mathcal{E}_m^{(\beta)}
[\rho^{(\alpha)}]|\varphi_{\text{in}}\rangle$ \cite{Ref20}, which
measures the overlap between the states
$|\varphi_{\text{in}}\rangle$ to be teleported and the output state
with the density operator
$\mathcal{E}_m^{(\beta)}[\rho^{(\alpha)}]$. Since
$|\varphi_{\text{in}}\rangle$ is in general unknown, it is more
beneficial to calculate the average fidelity (the fidelity averaged
over all possible Alice's measurement outcomes $m$ and all possible
pure input states $|\varphi_{\text{in}}\rangle$ on the Bloch sphere)
to quantify the teleportation process. This average fidelity is
defined as
\begin{equation}
 F^{(\beta)}[\rho^{(\alpha)}]=\frac{1}{4\pi}\int_0^{2\pi}d\phi
 \int_0^{\pi}d\theta\sin\theta\sum_{m=0}^{3}P_{m}^{(\alpha)}
 f_m^{(\beta)}[\rho^{(\alpha)}],
\end{equation}
where $4\pi$ is the solid angle.

\section{Teleportation with environment-disturbed recovery operations}
\label{sec:3}To begin with, we first explore the special case that
the channel state is prepared maximally and distributed perfectly
between Alice and Bob. For this case, we always have
$\rho^{(p)}=|{\rm{\Psi}}^0\rangle\langle{\rm{\Psi}}^0|$, combination
of this with equations (3) and (4) one can derive exactly the
complete analytical forms of the average fidelity with Bob's
recovery operation corrupted by the dissipative, noisy or dephasing
environment. Their explicit expressions are as follows
\begin{eqnarray}
 F^{(di)}[\rho^{(p)}]&=&\frac{1}{2}+\frac{1}{12}(e^{-\gamma t}-\alpha_1+2\alpha_2+\alpha_3)\nonumber\\&&
  +\frac{1}{6}e^{-\gamma t/2}\sin^{2}\frac{\omega t}{2}, \nonumber\\
 F^{(no)}[\rho^{(p)}]&=&\frac{7}{12}+\frac{1}{6}(\beta_2-\beta_1)+\frac{1}{12}e^{-2\gamma t}\nonumber\\&&
  +\frac{1}{6}e^{-\gamma t}\sin^{2}\frac{\omega t}{2}, \nonumber\\
 F^{(de)}[\rho^{(p)}]&=&\frac{2}{3}+\frac{1}{6}(\mu_2-\mu_1)+\frac{1}{6}e^{-\gamma t/2}\sin^{2}\frac{\omega t}{2}, \nonumber\\
\end{eqnarray}
where the corresponding parameters $\alpha_i$ ($i=1,2,3$), $\beta_j$
and $\mu_j$ ($j=1,2$) are given by
\begin{eqnarray}
 \alpha_1&=&\frac{4u(\gamma^2+\omega^2)\cosh(ut)-\gamma(\gamma^2+5\omega^2)\sinh(ut)}
            {4u(\gamma^2+2\omega^2)} \nonumber\\&&
            \times e^{-3\gamma t/4}+\frac{\omega^2}{\gamma^2+2\omega^2}, \nonumber\\
 \alpha_2&=&\frac{4ue^{-\gamma t/2}-[\gamma\sinh(ut)+4u\cosh(ut)]e^{-3\gamma t/4}}{8u}, \nonumber\\
 \alpha_3&=&\frac{4u\omega^2-[3\gamma\sinh(ut)+4u\cosh(ut)]\omega^2 e^{-3\gamma t/4}}{4u(\gamma^2+2\omega^2)}, \nonumber\\
 \beta_1&=&\frac{2v-[\gamma\sinh(vt)-2v\cosh(vt)]e^{-3\gamma t/2}}{4v}, \nonumber\\
 \beta_2&=&\frac{2ve^{-\gamma t}-[\gamma\sinh(vt)+2v\cosh(vt)]e^{-3\gamma t/2}}{4v}, \nonumber\\
 \mu_1&=&\frac{4u+[\gamma\sinh(ut)+4u\cosh(ut)]e^{-\gamma t/4}}{8u}, \nonumber\\
 \mu_2&=&\frac{4ue^{-\gamma t/2}+[\gamma\sinh(ut)-4u\cosh(ut)]e^{-\gamma t/4}}{8u}, \nonumber\\
\end{eqnarray}
with
\begin{equation}
 u=\frac{1}{4}\sqrt{\gamma^2-16\omega^2},~
 v=\frac{1}{2}\sqrt{\gamma^2-4\omega^2}.
\end{equation}
\begin{figure}
\centering
\resizebox{0.45\textwidth}{!}{%
\includegraphics{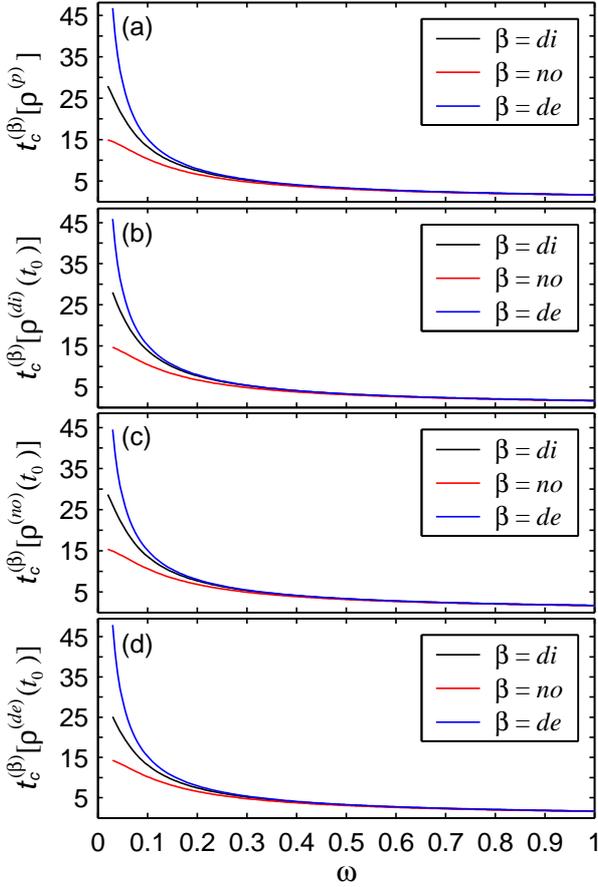}}
\caption{(Color online) Critical time
$t_c^{(\beta)}[\rho^{(\alpha)}]$ versus the rotation rate $\omega$,
where the decoherence rate is given by $\gamma=0.1$, and in (b), (c)
and (d) the curves are plotted with the transmission time $t_0=2$.}
\label{fig:2}
\end{figure}

\begin{figure}
\centering
\resizebox{0.45\textwidth}{!}{%
\includegraphics{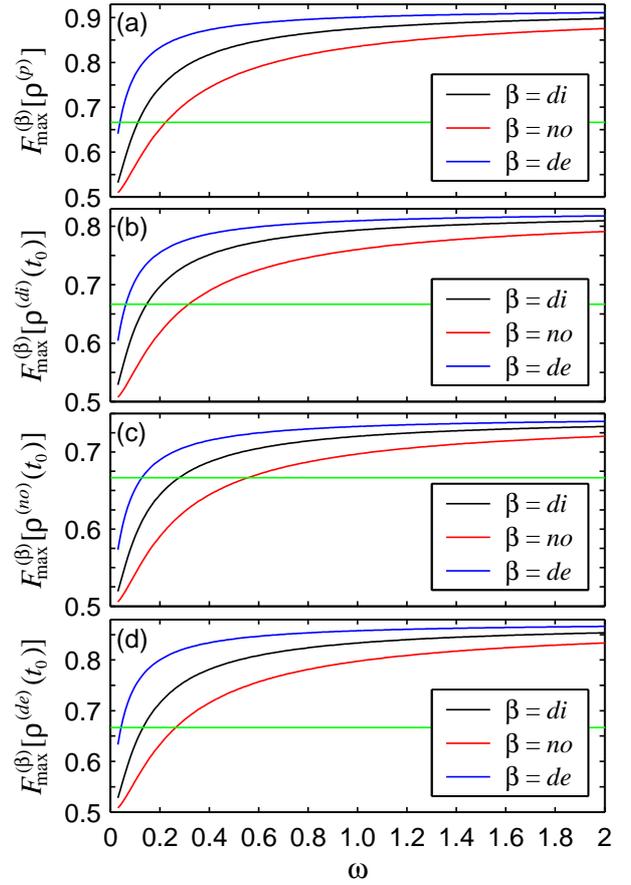}}
\caption{(Color online) The maximum of the average teleportation
fidelity $F_{\text{max}}^{(\beta)}[\rho^{(\alpha)}]$ versus the
rotation rate $\omega$, where the decoherence rate is given by
$\gamma=0.1$, and in (b), (c) and (d) the curves are plotted with
the transmission time $t_0=2$.} \label{fig:3}
\end{figure}

Since $F^{(\beta)}[\rho^{(p)}]$ ($\beta=di,no,de$) is a function of
the rotation rate $\omega$, decoherence rate $\gamma$ as well as the
time interval $t$ during which Bob performs the recovery operations,
there exists a critical time $t_c^{(\beta)}[\rho^{(p)}]$ at which
the average teleportation fidelity attains its maximum, denoted by
$F_{\text{max}}^{(\beta)}[\rho^{(p)}]$. This maximum is reached
whenever $t$ satisfies the relations
$\partial{F^{(\beta)}[\rho^{(p)}]}/\partial{t}=0$ and
$\partial^{2}{F^{(\beta)}[\rho^{(p)}]}/\partial{t^2}<0$. Complete
forms of the above two nonlinear equations can be derived
straightforwardly from equations (5), (6) and (7), however, since
the expressions of $F^{(\beta)}[\rho^{(p)}]$ are so complicated, it
is very difficult to obtain analytical solutions of them, thus we
resort to numerical methods. The corresponding results are plotted
in Figure 2(a) and Figure 3(a), from which one can see that for a
given decoherence rate $\gamma$, the critical time
$t_c^{(\beta)}[\rho^{(p)}]$ decreases monotonously with increasing
$\omega$ and we always have
$t_c^{(de)}[\rho^{(p)}]>t_c^{(di)}[\rho^{(p)}]\\>t_c^{(no)}[\rho^{(p)}]$,
while the maximum average fidelity
$F_{\text{max}}^{(\beta)}[\rho^{(p)}]$ increases with increasing
$\omega$ or decreasing $t_c^{(\beta)}[\rho^{(p)}]$ and approaches to
its asymptotic value when $\omega$ is infinitely large (when
$\gamma=0.1$ and $\omega=200$, we can obtain
$F_{\text{max}}^{(di)}[\rho^{(p)}]\simeq 0.92163$,
$F_{\text{max}}^{(no)}[\rho^{(p)}]\simeq 0.92138$ and
$F_{\text{max}}^{(de)}[\rho^{(p)}]\simeq 0.92177$, and the
differences between them becomes smaller and smaller with increasing
$\omega$). However, in contrast to those in reference \cite{Ref18}
(i.e., the channel is perfect while Bob's recovery operations are
corrupted by intrinsic, bit-flip or bit-phase-flip noise) where the
maximum of the average teleportation fidelity is always larger than
the classical limiting value of $2/3$ \cite{Ref21}, there exists a
critical rotation rate $\omega_c^{(\beta)}[\rho^{(p)}]$ smaller than
which the standard teleportation protocol $\mathcal{P}_0$ will fail
to achieve nonclassical fidelity. $\omega_c^{(\beta)}[\rho^{(p)}]$
increases monotonously with increasing value of $\gamma$, and when
$\gamma=0.1$ we have $\omega_c^{(di)}[\rho^{(p)}]\simeq 0.11192$,
$\omega_c^{(no)}[\rho^{(p)}]\simeq 0.12999$ and
$\omega_c^{(de)}[\rho^{(p)}]\simeq 0.03829$. Moreover, for any fixed
decoherence rate $\gamma$, we always have the relations
$F^{(de)}[\rho^{(p)}]>F^{(di)}[\rho^{(p)}]>F^{(no)}[\rho^{(p)}]$
associated with the maximum average teleportation fidelity, which
indicates that the devastating effects of the noisy environment is
more severe than those of the dissipative and the dephasing
environments.

Consider now the situation in which the transmission process of the
two qubits constitute the quantum channel are infected with the
dissipative environment (see Figure 1) under an interval of time,
say $t_0$. Then the initial maximally entangled Bell state
$|\rm{\Psi}^0\rangle$ will be destroyed, and by solving the
appropriate master equation (1) with $|\rm{\Psi}^0\rangle$ as the
initial condition and the generators of decoherence given by
$\mathcal {L}_k=\sigma_k^{-}$, one can obtain the analytical
expressions of the nonzero elements of $\rho^{(di)}(t_0)$ explicitly
as $\rho_{11}^{(di)}(t_0)=e^{-2\gamma t_0}/2$,
$\rho_{14,41}^{(di)}(t_0)=e^{-\gamma t_0}/2$,
$\rho_{22,33}^{(di)}(t_0)=(e^{-\gamma t_0}-e^{-2\gamma t_0})/2$, and
$\rho_{44}^{(di)}(t_0)=1-e^{-\gamma t_0}+e^{-2\gamma t_0}/2$, whose
concurrence (a measure of pairwise entanglement introduced by
Wootters \cite{Ref22}) $C(t_0)=e^{-2\gamma t_0}$ decays
exponentially with increasing time $t_0$ , which forecasts the
possible depression of the teleportation fidelity of the expected
outcomes. By solving again the master equation (1) with
$\rho^{(di)}(t_0)$ as the initial state and
$\hat{H}_{m}=-\omega\sigma^{m}/2$ as the Hamiltonian of the system
and then combining the corresponding solutions with equations (3)
and (4), one can derive the explicit forms of the average fidelity
$F^{(\beta)}[\rho^{(di)}]$ with Bob's recovery operations disturbed
by the dissipative, noisy and dephasing environments, as
\begin{eqnarray}
 F^{(di)}[\rho^{(di)}]&=&\frac{1}{2}+\frac{1}{12}(2e^{-2\gamma t_0}-e^{-\gamma t_0})\nonumber\\&&
                         \times(e^{-\gamma t}-\alpha_1+\alpha_3)+\frac{1}{6}\alpha_3\nonumber\\&&
                         +\frac{1}{6}e^{-\gamma t_0}\left(\alpha_2-\alpha_3+e^{-\gamma t/2}\sin^{2}\frac{\omega t}{2}\right), \nonumber\\
 F^{(no)}[\rho^{(di)}]&=&\frac{2}{3}+\frac{1}{6}(e^{-2\gamma t_0}-e^{-\gamma t_0})(e^{-2\gamma t}-2\beta_1+1)\nonumber\\&&
                         -\frac{1}{6}\beta_1+\frac{1}{6}e^{-\gamma t_0}\left(\beta_2+e^{-\gamma t}\sin^{2}\frac{\omega t}{2}\right), \nonumber\\
 F^{(de)}[\rho^{(di)}]&=&\frac{2}{3}+\frac{1}{3}(e^{-2\gamma t_0}-e^{-\gamma t_0})(1-\mu_1)-\frac{1}{6}\mu_1\nonumber\\&&
                         +\frac{1}{6}e^{-\gamma t_0}\left(\mu_2+e^{-\gamma t/2}\sin^{2}\frac{\omega t}{2}\right), \nonumber\\
\end{eqnarray}
where the corresponding parameters $\alpha_i$ ($i=1,2,3$), $\beta_j$
and $\mu_j$ ($j=1,2$) appeared in the above equations are completely
the same as those expressed in equation (6). Since their expressions
are still so complicated, we resort to numerical methods again. The
critical time $t_c^{(\beta)}[\rho^{(di)}(t_0)]$ at which the average
fidelity attains its maximum versus the rotation rate $\omega$ are
displayed in Figure 2(b), while the maximum of
$F^{(\beta)}[\rho^{(di)}(t_0)]$, denoted by
$F_{\text{max}}^{(\beta)}[\rho^{(di)}(t_0)]$, versus $\omega$ are
displayed in Figure 3(b), both with the decoherence rate and the
transmission time given by $\gamma=0.1$ and $t_0=2$, respectively.
From Figure 2(b) one can see that although the dissipative
environment disentangling the two qubits involved in the quantum
channel exponentially, the $\omega$ dependence of the critical time
$t_c^{(\beta)}[\rho^{(di)}(t_0)]$ displays nearly the same behaviors
as those of $t_c^{(\beta)}[\rho^{(p)}]$. In general, we have
$t_c^{(\beta)}[\rho^{(di)}(t_0)]>t_c^{(\beta)}[\rho^{(p)}]$,
$t_c^{(\beta)}[\rho^{(no)}(t_0)]>t_c^{(\beta)}[\rho^{(p)}]$ and
$t_c^{(\beta)}[\rho^{(de)}(t_0)]<t_c^{(\beta)}[\rho^{(p)}]$,
however, the differences between them are very small. Particularly,
in the large $\omega$ region this difference can even be neglected
(cf. the curves displayed in Figures 2(a) and 2(b)). When
considering the maximum average fidelity
$F_{\text{max}}^{(\beta)}[\rho^{(di)}(t_0)]$, as can be seen from
Figure 3(b), it also increases with increasing $\omega$, and
approaches to its asymptotic value which is smaller than that of
Alice and Bob share the ideal channel state $|\rm{\Psi}^0\rangle$ in
the limit of $\omega\rightarrow\infty$ (when $\gamma=0.1$, $t_0=2$
and $\omega=200$, one can obtain
$F_{\text{max}}^{(di)}[\rho^{(di)}(t_0)]\simeq 0.82629$,
$F_{\text{max}}^{(no)}[\rho^{(di)}(t_0)]\simeq 0.82609$ and
$F_{\text{max}}^{(de)}[\rho^{(di)}(t_0)]\simeq 0.82638$). The
depression of the average fidelity can be attributed to the
exponential decay of the entanglement of the channel state during
the transmission time $t_0$.
\begin{figure}
\centering
\resizebox{0.45\textwidth}{!}{%
\includegraphics{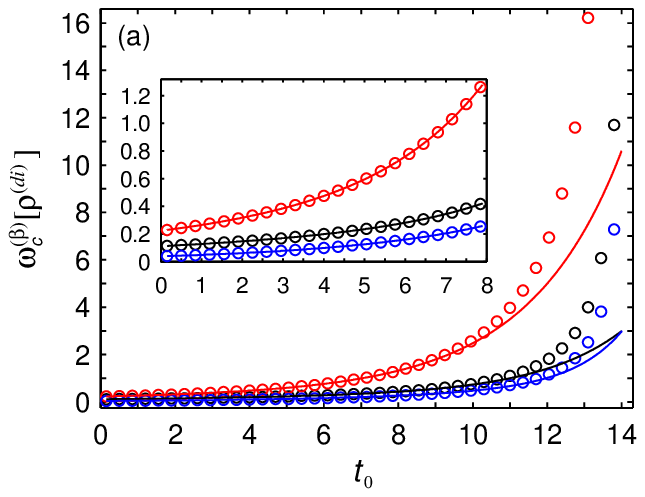}}
\resizebox{0.45\textwidth}{!}{%
\includegraphics{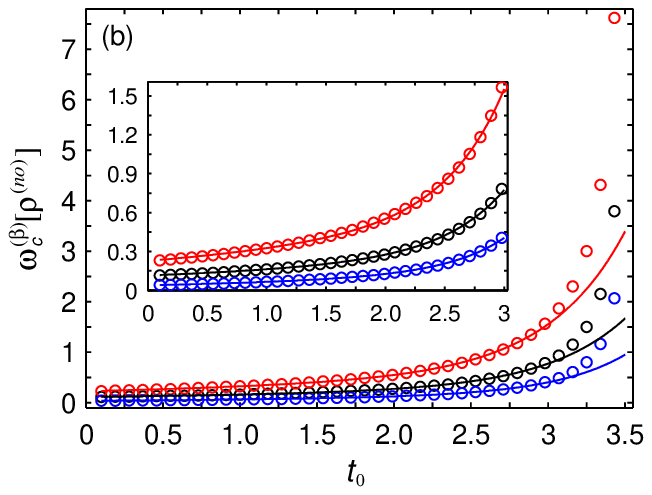}}
\resizebox{0.45\textwidth}{!}{%
\includegraphics{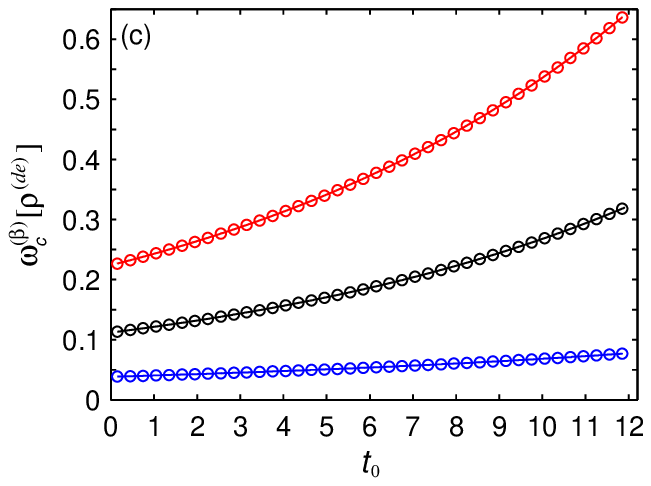}}
\caption{(Color online) Critical rotation rate
$\omega_c^{(\beta)}[\rho^{(\alpha)}]$ versus the transmission time
$t_0$ with the decoherence rate given by $\gamma=0.1$ and Bob's
recovery operations corrupted by the dissipative (denoted by the
black circles with $t_0\in[0.15,13.8]$), noisy (denoted by the red
circles with $t_0\in[0.1,3.43]$) and dephasing (denoted by the blue
circles with $t_0\in[0.15,11.85]$) environment. Here the black, red
and blue lines show the corresponding data fitting results with
$t_0\in[0.15,7.85]$ (a), $t_0\in[0.1,2.98]$ (b), and
$t_0\in[0.15,11.85]$ (c). Moreover, the insets in (a) and (b) show
behaviors of $\omega_c^{(\beta)}[\rho^{(\alpha)}]$ in the small
regions of $t_0$, where the labels of the horizontal and vertical
axes are omitted for compaction of the plots.} \label{fig:3}
\end{figure}

\begin{table}[!h]
\tabcolsep 0pt \caption{The coefficients $a$, $b$, $c$ and $d$
appeared in the data fitting equation
$\omega_c^{(\beta)}[\rho^{(\alpha)}]=ae^{bt_0}+ce^{dt_0}$ with
different imperfect channel states (denoted by $\alpha=di$, $no$ or
$de$) and different environment-disturbed recovery operations
(denoted by $\beta=di$, $no$ or $de$).} \vspace*{-8pt}
\begin{center}
\def\temptablewidth{0.475\textwidth}
{\rule{\temptablewidth}{1pt}}
\begin{tabular*}{\temptablewidth}{@{\extracolsep{\fill}}ccllll}
    $\alpha$ & $\beta$  & \it a    & \it b       & \it c       & \it d   \\  \hline
             & \it di   & 0.1087   & 0.1224      & 0.003274    & 0.4707  \\
    \it di   & \it no   & 0.1981   & 0.1317      & 0.02590     & 0.4206  \\
             & \it de   & 0.03886  & 0.2333      & 0.00001795  & 0.8284  \\
             & \it di   & 0.1129   & 0.3115      & 0.001157    & 2.014   \\
    \it no   & \it no   & 0.2251   & 0.3220      & 0.001967    & 2.064   \\
             & \it de   & 0.03893  & 0.4788      & 0.0002784   & 2.254   \\
             & \it di   & 0.02222  & -0.0006913  & 0.08975     & 0.1007  \\
    \it de   & \it no   & 0.05079  & 0.008443    & 0.1731      & 0.1021  \\
             & \it de   & 0.03666  & 0.04937     & 0.001702    & 0.1574
\end{tabular*}
{\rule{\temptablewidth}{1pt}}
\end{center}
\end{table}

For any fixed decoherence rate $\gamma$, equation (8) also gives the
constraint on a critical rotation rate
$\omega_c^{(\beta)}[\rho^{(di)}]$ smaller than which the standard
teleportation protocol will fail to yield an average fidelity better
than classically possible. $\omega_c^{(\beta)}[\rho^{(di)}]$
increases with increasing value of $\gamma$ (when $\gamma=0.1$ and
$t_0=2$, we can obtain $\omega_c^{(di)}[\rho^{(di)}]\simeq0.14719$,
$\omega_c^{(no)}[\rho^{(di)}]\simeq 0.31763$ and
$\omega_c^{(de)}[\rho^{(di)}]\simeq 0.06229$). However, in big
contrast to those with the ideal channel state $|\rm{\Psi}^0\rangle$
in which Bob's imperfect recovery operations do not result in an
average fidelity worse than that achievable via purely classical
communication alone, our numerical results demonstrated that here
$F_{\text{max}}^{(\beta)}[\rho^{(di)}]$ cannot exceed its classical
limiting value of $2/3$ when the transmission time $t_0$ is longer
than a critical value $t_{0,c}$ (for $\gamma=0.1$ this critical
value can be obtained approximately as $t_{0,c}\simeq 14.212$). In
the region of $t_0<t_{0,c}$, we plotted the critical rotation rate
$\omega_c^{(\beta)}[\rho^{(di)}]$ versus the transmission time $t_0$
in Figure 4(a) as hollow circles, from which one can see that
$\omega_c^{(\beta)}[\rho^{(di)}]$ increases with increasing time
$t_0$. Particularly, in the small region of $t_0$,
$\omega_c^{(\beta)}[\rho^{(di)}]$ can be approximately fitted as
\begin{equation}
 \omega_c^{(\beta)}[\rho^{(di)}]=ae^{bt_0}+ce^{dt_0},
\end{equation}
where the coefficients $a$, $b$, $c$ and $d$ are given in Table 1.
The corresponding data fitting results are plotted in Figure 4(a) as
black, red and blue lines, respectively. Clearly, in the small $t_0$
region, $\omega_c^{(\beta)}[\rho^{(di)}]$ increases exponentially
with the increase of $t_0$ and the data fitting results are very
well, while in the large $t_0$ region, the divergence of the data
fitting results from the real critical rotation rate becomes larger
and larger with increasing $t_0$.

If the transmission process is infected with the noisy environment,
suppose it takes time $t_0$ for the two qubits to reach Alice and
Bob. Then by solving the appropriate master equation (1) with the
generators of decoherence given by $\mathcal {L}_{k,1}=\sigma_k^{-}$
and $\mathcal {L}_{k,2}=\sigma_k^{+}$, we can obtain explicitly the
nonzero elements of $\rho^{(no)}(t_0)$ as
$\rho_{11,44}^{(no)}(t_0)=(1+e^{-4\gamma t_0})/4$,
$\rho_{14,41}^{(no)}(t_0)=e^{-2\gamma t_0}/2$ and
$\rho_{22,33}^{(no)}(t_0)=(1-e^{-4\gamma t_0})/4$. The concurrence
[22] of $\rho^{(no)}(t_0)$ is given by $C(t_0)={\text{max}}\{0,
e^{-2\gamma t_0}+e^{-4\gamma t_0}/2-1/2\}$. Different from the
former case, here the concurrence decays with increasing $t_0$ and
becomes zero abruptly at $\gamma t_0=[\ln(\sqrt{2}+1)]/2\simeq
0.440687$, and remains zero thereafter, which is known as
entanglement sudden death (ESD) observed previously by Yu and Eberly
\cite{Ref23} and has been extensively studied recently. Thus it is
reasonable to conjecture that when the transmission time $t_0$
becomes longer than $[\ln(\sqrt{2}+1)]/2\gamma$, the teleportation
protocol will fail to achieve nonclacssical fidelity. In fact, the
critical transmission time for
$F_{\text{max}}^{(\beta)}[\rho^{(no)}({t_0})]$ is much smaller than
$[\ln(\sqrt{2}+1)]/2\gamma$ (see the following text). To see this
more clearly, we solve again the master equation (1) with
$\rho^{(no)}(t_0)$ as the initial state and
$\hat{H}_m=-\omega\sigma^{m}/2$ as the system Hamiltonian, and by
combination of the corresponding solutions with equations (3) and
(4), we obtain
\begin{eqnarray}
 F^{(di)}[\rho^{(no)}]&=&\frac{1}{2}+\frac{1}{12}e^{-4\gamma t_0}(e^{-\gamma t}-\alpha_1+\alpha_3)\nonumber\\&&
                         +\frac{1}{6}e^{-2\gamma t_0}\left(\alpha_2+e^{-\gamma t/2}\sin^2\frac{\omega t}{2}\right), \nonumber\\
 F^{(no)}[\rho^{(no)}]&=&\frac{1}{2}+\frac{1}{12}e^{-4\gamma t_0}(e^{-2\gamma t}-2\beta_1+1)\nonumber\\&&
                         +\frac{1}{6}e^{-2\gamma t_0}\left(\beta_2+e^{-\gamma t}\sin^2\frac{\omega t}{2}\right), \nonumber\\
 F^{(de)}[\rho^{(no)}]&=&\frac{1}{2}+\frac{1}{6}e^{-4\gamma t_0}(1-\mu_1)\nonumber\\&&
                         +\frac{1}{6}e^{-2\gamma t_0}\left(\mu_2+e^{-\gamma t/2}\sin^2\frac{\omega t}{2}\right). \nonumber\\
\end{eqnarray}

The maximum of the average fidelity
$F^{(\beta)}[\rho^{(no)}({t_0})]$ is achieved at the critical time
$t_{c}^{(\beta)}[\rho^{(no)}({t_0})]$, whose $\omega$ dependence is
plotted in Figure 2(c) with the decoherence rate and the
transmission time given by $\gamma=0.1$ and $t_0=2$, while the
$\omega$ dependence of
$F_{\text{max}}^{(\beta)}[\rho^{(no)}({t_0})]$ is plotted in Figure
3(c), still with $\gamma=0.1$ and $t_0=2$. From these two figures
one can observe that they display very similar behaviors as those
with the transmission process of the two qubits disturbed by the
dissipative environment, i.e., $t_{c}^{(\beta)}[\rho^{(no)}({t_0})]$
decreases while $F_{\text{max}}^{(\beta)}[\rho^{(no)}({t_0})]$
increases with increasing value of $\omega$. But now the magnitudes
of $F_{\text{max}}^{(\beta)}[\rho^{(no)}({t_0})]$ are further
depressed (for the case of $\gamma=0.1$, $t_0=2$ and $\omega=200$,
we have $F_{\text{max}}^{(di)}[\rho^{(no)}({t_0})]\simeq 0.74651$,
$F_{\text{max}}^{(no)}[\rho^{(no)}({t_0})]\simeq 0.74637$ and
$F_{\text{max}}^{(de)}[\rho^{(no)}({t_0})]\simeq 0.74658$). This
phenomenon may be caused by the competition between the two
generators $\mathcal {L}_{k,1}=\sigma_k^{-}$ and $\mathcal
{L}_{k,2}=\sigma_k^{+}$ of decoherence. Also we found that the
critical rotation rate $\omega_{c}^{(\beta)}[\rho^{(no)}({t_0})]$
after which $F_{\text{max}}^{(\beta)}[\rho^{(no)}({t_0})]$ exceeds
the classical limiting value $2/3$ is enhanced. For instance, we
have $\omega_{c}^{(di)}[\rho^{(no)}({t_0})]\simeq 0.27823$,
$\omega_{c}^{(no)}[\rho^{(no)}({t_0})]\simeq 0.55646$ and
$\omega_{c}^{(de)}[\rho^{(no)}({t_0})]\simeq 0.12794$ if
$\gamma=0.1$ and $t_0=2$. Moreover, the maximum average
teleportation fidelity cannot exceed $2/3$ when the transmission
time $t_0$ is larger than a critical value $t_{0,c}$ (for
$\gamma=0.1$ this critical value can be approximated as
$t_{0,c}\simeq3.549$). When $t_0<t_{0,c}$, from Figure 4(b) one can
observe that $\omega_{c}^{(\beta)}[\rho^{(no)}]$ can also be fitted
very well as $\omega_c^{(\beta)}[\rho^{(no)}]=ae^{bt_0}+ce^{dt_0}$
(with the coefficients $a$, $b$, $c$ and $d$ given in Table 1) in
the small $t_0$ region, while in the large $t_0$ region, the
divergence of the corresponding data fitting results from the real
critical rotation rate also increases with increasing $t_0$.

For the case that the transmission process of the two qubits from
the EPR source to Alice and Bob is infected with the dephasing
environment, we obtain the nonzero elements of $\rho^{(de)}(t_0)$ as
$\rho_{11,44}^{(de)}(t_0)=1/2$ and
$\rho_{14,41}^{(de)}(t_0)=e^{-\gamma t_0}/2$, which yields the
concurrence [22] $C(t_0)=e^{-\gamma t_0}$. $C(t_0)$ also decays
exponentially with increasing $t_0$, with however, the decay rate
smaller than that under the influence of the dissipative
environment, thus it is natural to expect an enhancement of the
average teleportation fidelity. By solving the master equation (1)
with $\rho^{(de)}(t_0)$ as the initial state and
$\hat{H}_m=-\omega\sigma^{m}/2$ as the system Hamiltonian, and then
inserting the corresponding solutions into equations (3) and (4), we
obtain $F^{(\beta)}[\rho^{(de)}]$ with Bob's recovery operations
executed in the presence of the dissipative, noisy and dephasing
environments, as
\begin{eqnarray}
 F^{(di)}[\rho^{(de)}]&=&\frac{1}{2}+\frac{1}{12}(e^{-\gamma t}-\alpha_1+\alpha_3)\nonumber\\&&
                         +\frac{1}{6}e^{-\gamma t_0}\left(\alpha_2+e^{-\gamma t/2}\sin^2\frac{\omega t}{2}\right), \nonumber\\
 F^{(no)}[\rho^{(de)}]&=&\frac{7}{12}+\frac{1}{12}e^{-2\gamma t}-\frac{1}{6}\beta_1\nonumber\\&&
                         +\frac{1}{6}e^{-\gamma t_0}\left(\beta_2+e^{-\gamma t}\sin^2\frac{\omega t}{2}\right), \nonumber\\
 F^{(de)}[\rho^{(de)}]&=&\frac{2}{3}-\frac{1}{6}\mu_1+\frac{1}{6}e^{-\gamma t_0}\left(\mu_2+e^{-\gamma t/2}\sin^2\frac{\omega t}{2}\right). \nonumber\\
\end{eqnarray}

Still there exists a critical time
$t_{c}^{(\beta)}[\rho^{(de)}({t_0})]$ at which the average fidelity
$F^{(\beta)}[\rho^{(de)}]$ attains its maximum value.
$t_{c}^{(\beta)}[\rho^{(de)}({t_0})]$ and
$F_{\text{max}}^{(\beta)}[\rho^{(de)}({t_0})]$ versus the rotation
rate $\omega$ are shown in Figure 2(d) and Figure 3(d), where the
decoherence rate and the transmission time are chosen to be
$\gamma=0.1$ and $t_0=2$, respectively. These two figures display
very similar behaviors with those of the former cases. But now the
magnitudes of $F_{\text{max}}^{(\beta)}[\rho^{(de)}({t_0})]$ are
slightly enhanced (e.g., when $\gamma=0.1$, $t_0=2$ and
$\omega=200$, we have
$F_{\text{max}}^{(di)}[\rho^{(de)}({t_0})]\simeq 0.87496$,
$F_{\text{max}}^{(no)}[\rho^{(de)}({t_0})]\simeq 0.87474$ and
$F_{\text{max}}^{(de)}[\rho^{(de)}({t_0})]\simeq 0.87510$) compared
with that of the transmission process being disturbed by the noisy
environment. Also our numerical results demonstrated that in order
for $F_{\text{max}}^{(\beta)}[\rho^{(de)}({t_0})]$ to exceed $2/3$,
the transmission time must be shorter than a critical value
$t_{0,c}$ ($t_{0,c}\simeq12.194$ when $\gamma=0.1$). In the region
of $t_0<t_{0,c}$, the critical rotation rate
$\omega_c^{(\beta)}[\rho^{(de)}]$ after which
$F_{\text{max}}^{(\beta)}[\rho^{(de)}({t_0})]$ exceeds $2/3$
exhibits an exponential increase with increasing $t_0$ (we have
$\omega_c^{(di)}[\rho^{(de)}]\simeq 0.13194$,
$\omega_c^{(no)}[\rho^{(de)}]\simeq 0.26389$ and
$\omega_c^{(de)}[\rho^{(de)}]\simeq 0.04273$ when $\gamma=0.1$ and
$t_0=2$ ), and as can be seen from Figure 4(c), it can also be
fitted very well as $\omega_c^{(\beta)}[\rho^{(de)}]=ae^{b
t_0}+ce^{d t_0}$, where the coefficients $a$, $b$, $c$ and $d$ are
displayed in Table 1.

Now we make some discussion about the physical implications of the
phenomena displayed in Figures 2 and 3. Since decoherence occurs in
all real physical entities, it is essential to reduce the gap
between the experimentally achieved coherence lifetimes and those
required by theory for quantum teleportation to become practical.
For spin qubits the coherence times in general are very short (e.g.,
coherence time of about 3.0 $\mu\text{s}$ has been reported for
quantum dot electron spins \cite{Ref24}, and conherence time of
about 30 $\mu\text{s}$ has been demonstrated experimentally in
irradiated malonic acid crystals at temperature 50 K \cite{Ref25}),
thus for the teleportation protocol to remain effective, the
rotation operation to the target qubit must be performed on a
timescale in which the coherence of the spin is preserved. The
curves in Figures 2 and 3 show evidently that the quality of quantum
teleportation can be improved significantly by increasing the
rotation rate $\omega$ and thus the resultant gate operation times
can be shortened, this indicates that the large rotation rate to the
target spin is beneficial to quantum teleportation.

Finally, we would like to make a comparison of the robustness of the
standard teleportation protocol $\mathcal {P}_0$ when it is executed
with different (perfect or decohering) quantum channels as well as
different environment-disturbed recovery operations. First, for
fixed $\alpha=p$, $di$, $no$ or $de$ (i.e., the case with the same
quantum channels but different imperfect recovery operations), we
always have $F_{\text{max}}^{(de)}[\rho^{(\alpha)}({t_0})]>
F_{\text{max}}^{(di)}[\rho^{(\alpha)}({t_0})]>
F_{\text{max}}^{(no)}[\rho^{(\alpha)}({t_0})]$. This relation
reveals that under the condition of the same decoherence rate
$\gamma$, the destructive effects of the noisy environment on Bob's
recovery operation is more severe than that of the dissipative or
the dephasing environment, thus it is reasonable to obtain the
relation $\omega_c^{(de)}[\rho^{(\alpha)}(t_0)]<
\omega_c^{(di)}[\rho^{(\alpha)}(t_0)]<
\omega_c^{(no)}[\rho^{(\alpha)}(t_0)]$ for the critical rotation
rate. Second, for any fixed $\beta=di$, $no$ or $de$ (i.e., the case
with different quantum channels but the same recovery operations),
we always have the following two relations:
$F_{\text{max}}^{(\beta)}[\rho^{(p)}]>
F_{\text{max}}^{(\beta)}[\rho^{(de)}({t_0})]>
F_{\text{max}}^{(\beta)}[\rho^{(di)}({t_0})]>
F_{\text{max}}^{(\beta)}[\rho^{(no)}({t_0})]$ for the maximum
average teleportation fidelity, and $\omega_c^{(\beta)}[\rho^{(p)}]<
\omega_c^{(\beta)}[\rho^{(de)}(t_0)]<
\omega_c^{(\beta)}[\rho^{(di)}(t_0)]<
\omega_c^{(\beta)}[\rho^{(no)}(t_0)]$ for the critical rotation
rate. These demonstrate again that the devastating effects imposed
by the noisy environment on the standard teleportation protocol
$\mathcal {P}_0$ is the most serious one compared with the other
cases.

\section{Summary and discussion}
\label{sec:4} In summary, we have studied standard teleportation
process of an arbitrary one-qubit state with both the transmission
process of the two qubits constitute the quantum channel and the
recovery operations performed by Bob disturbed by the dissipative,
noisy and dephasing environments. Through detailed analyzation of
the average fidelities with different situations, we demonstrated
that provided Alice and Bob share an ideal channel state, Bob's
environment-disturbed recovery operations do not eliminate the
possibility for teleporting the one-qubit state with nonclassical
fidelity. When the transmission process of the two qubits is
corrupted by decohering environment, however, there exists a
constraint on the transmission time $t_0$, denoted by $t_{0,c}$,
after which the standard teleportation protocol $\mathcal{P}_0$ will
fail to attain an average fidelity better than classically possible.

We have also compared the robustness of the teleportation protocol
$\mathcal{P}_0$ with different (perfect or decohered) quantum
channels and different imperfect recovery operations, and obtained
the following two general relations for the average fidelities: (i)
$F_{\text{max}}^{(de)}[\rho^{(\alpha)}(t_0)]>
F_{\text{max}}^{(di)}[\rho^{(\alpha)}(t_0)]>
F_{\text{max}}^{(no)}[\rho^{(\alpha)}(t_0)]$, and (ii)
$F_{\text{max}}^{(\beta)}[\rho^{(p)}]>
F_{\text{max}}^{(\beta)}[\rho^{(de)}(t_0)]>
F_{\text{max}}^{(\beta)}[\rho^{(di)}(t_0)]
>F_{\text{max}}^{(\beta)}[\rho^{(no)}(t_0)]$.
We have therefore revealed that under the condition of the same
decoherence rate $\gamma$, the standard teleportation protocol
$\mathcal{P}_0$ is significantly more fragile under the influence of
the noisy environment than those under the influence of the
dissipative and the dephasing environments.

Besides decoherence processes considered in this work, Alice's Bell
state measurement may also be disturbed by the decohering
environments \cite{Ref14}. Even though not shown here it will result
in a further depression of the average teleportation fidelity. Thus
in spite of the existence of certain special conditions under which
the teleportation fidelity may be enhanced to some extent by the
local environment \cite{Ref26,Ref27}, finding ways to stabilize the
entangled channel state and to minimize, delay or even eliminate the
devastating effects of the decohering environment is still an
challenging task in the physical realization of the teleportation
protocol. Moreover, understanding the influence of surrounding
environments on an open quantum system such as the mechanisms of
decoherence and the dynamics of entanglement is also both of
fundamental interest in quantum foundation issues and of practical
importance in quantum information theory.
\\
\\
\\
This work was supported in part by the NSF of Shaanxi Province under
grant Nos. 2010JM1011 and 2009JQ8006, the Specialized Research
Program of Education Department of Shaanxi Provincial Government
under grant Nos. 2010JK843 and 2010JK828, and the Youth Foundation
of XUPT under Grant No. ZL2010-32.


\begin{thebibliography}{27}
\bibitem{Ref1}C.H. Bennett, G. Brassard, C. Cr\'{e}peau, R. Jozsa, A. Peres, W.K. Wootters, \PRL \textbf{70}, 1895 (1993)
\bibitem{Ref2}J. Lee, M.S. Kim, \PRL \textbf{84}, 4236 (2000)
\bibitem{Ref3}G. Bowen, S. Bose, \PRL \textbf{87}, 267901 (2001)
\bibitem{Ref4}F. Verstraete, H. Verschelde, \PRL \textbf{90}, 097901 (2003)
\bibitem{Ref5}Y. Yeo, W.K. Chua, \PRL \textbf{96}, 060502 (2006)
\bibitem{Ref6}D. Bouwmeester, J.W. Pan, K. Mattle, M. Eibl, H. Weinfurter, A. Zeilinger, Nature \textbf{390}, 575 (1997)
\bibitem{Ref7}S. Olmschenk, D.N. Matsukevich, P. Maunz, D. Hayes, L.M. Duan, C. Monroe, Science \textbf{323}, 486 (2009)
\bibitem{Ref8}M. Riebe, H. H\"{a}ffner, C.F. Roos et al, Nature \textbf{429}, 734 (2004)
\bibitem{Ref9}M.D. Barrett, J. Chiaverini, T. Schaetz et al, Nature \textbf{429}, 737 (2004).
\bibitem{Ref10}A.R.R. Carvalho, F. Mintert, A. Buchleitner, \PRL \textbf{93}, 230501 (2004)
\bibitem{Ref11}J. Wang, H. Batelaan, J. Podany, A.F. Starace, J. Phys. B: At. Mol. Opt. Phys. \textbf{39}, 4343 (2006)
\bibitem{Ref12}A. Abliz, H.J. Gao, X.C. Xie, Y.S. Wu, W.M. Liu, \PRA \textbf{74}, 052105 (2006)
\bibitem{Ref13}N. Buri\'{c}, \PRA \textbf{77}, 012321 (2008)
\bibitem{Ref14}S. Oh, S. Lee, H.W. Lee, \PRA \textbf{66}, 022316 (2002)
\bibitem{Ref15}E. Jung, M.R. Hwang, D.K. Park, J.W. Son, S. Tamaryan, J. Phys. A: Math. Theor. \textbf{41}, 385302 (2008)
\bibitem{Ref16}E. Jung, M.R. Hwang, Y.H. Ju et al, \PRA \textbf{78}, 012312 (2008)
\bibitem{Ref17}Rao D.D. Bhaktavatsala, P.K. Panigrahi, C. Mitra, \PRA \textbf{78}, 022336 (2008)
\bibitem{Ref18}Y. Yeo, Z.W. Kho, L. Wang, EPL \textbf{86}, 40009 (2009)
\bibitem{Ref19}H.P. Breuer, F. Petruccione, {\it The Theory of Open Quantum Systems} (Oxford University Press, Oxford, 2002)
\bibitem{Ref20}M.A. Nielsen, I.L. Chuang, {\it Quantum Computation and Quantum Information} (Cambridge University Press, Cambridge, 2000)
\bibitem{Ref21}M. Horodecki, P. Horodecki, R. Horodecki, \PRA \textbf{60}, 1888 (1999)
\bibitem{Ref22}W.K. Wootters, \PRL \textbf{80}, 2245 (1998)
\bibitem{Ref23}T. Yu, J.H. Eberly, \PRL \textbf{93}, 140404 (2004)
\bibitem{Ref24}A. Greilich, D.R. Yakovlev, A. Shabaev et al., Science \textbf{313}, 341 (2006)
\bibitem{Ref25}J.F. Du, X. Rong, N. Zhao, Y. Wang, J.H. Yang, R.B. Liu, Nature \textbf{461}, 1265 (2009)
\bibitem{Ref26}P. Badziag, M. Horodecki, P. Horodecki, R. Horodecki, \PRA \textbf{62}, 012311 (2000)
\bibitem{Ref27}Y. Yeo, \PRA \textbf{78}, 022334 (2008)
\end{thebibliography}
%
\newcommand{\PRL}{Phys. Rev. Lett. }
\newcommand{\PRA}{Phys. Rev. A }
%

\end{document}